\documentclass{tlp}

\usepackage{xypic}
\usepackage[arrow,matrix,frame,curve,ps]{xy}

\newcommand{\chr}{{CHR}}

\newcommand{\false}{{\it false}}

\def\bq{\begin{quote}}
\def\eq{\end{quote}}

\newcommand{\comment}[1]{}
\newcommand{\kw}[1]{{\bf #1}} 

\newcommand{\newterm}[1]{{\em #1}}
\newcommand{\propag}{\ensuremath{\Rightarrow}}

\newcommand{\lhs}{{lhs}}
\newcommand{\rhs}{{rhs}}

\newcommand{\baselhs}{\ensuremath{Base_\lhs}}
\newcommand{\candlhs}{\ensuremath{Cand_\lhs}}
\newcommand{\candrhs}{\ensuremath{Cand_\rhs}}

\newcommand{\ctheory}{\ensuremath{{CT}}}

\newcommand{\Llhs}{\ensuremath{{L}}} 
\newcommand{\newruleset}{\ensuremath{{\cal R}}}

\newtheorem{definition}{Definition}[section]
\newtheorem{theorem}{Theorem}[section]
\newtheorem{example}{Example}[section]

\submitted{31 August 2002}
\revised{29 February 2004, 30 June 2004}
\accepted{9 August 2004}

\begin{document}

\title[Generation of {CHR} Constraint Solvers]
{Automatic Generation of {CHR} \\  Constraint Solvers}
\author[S. Abdennadher and C. Rigotti]
{SLIM ABDENNADHER\\
Computer Science Department, German University In Cairo, Egypt\\
  \email{Slim.Abdennadher@guc.edu.eg}
\and
 CHRISTOPHE RIGOTTI\\
LIRIS Laboratory, INSA Lyon, France\\
\email{Christophe.Rigotti@insa-lyon.fr}}

\pagerange{\pageref{firstpage}--\pageref{lastpage}}
\volume{\textbf{10} (3):}
\jdate{March 2000}
\setcounter{page}{1}
\pubyear{2000}

\maketitle
\label{firstpage}

\begin{abstract}
  
  In this paper, we present a framework for automatic generation of
  CHR solvers given the logical specification of the constraints.
  This approach takes advantage of the power of tabled resolution for
  constraint logic programming, in order to check the validity of the
  rules. Compared to previous
  work~\cite{AptMonfroy99,RingeissenMonfroy99,AbdennadherRigotti00,AbdennadherRigotti01b},
  where different methods for automatic generation of constraint
  solvers have been proposed, our approach enables the generation of
  more expressive rules (even recursive and splitting rules) that can
  be used directly as CHR solvers.

\end{abstract}


\section{Introduction}

Constraint Handling Rules (\chr)~\cite{Fru98} is a high-level language
especially designed for writing constraint solvers.  CHR is
essentially a committed-choice language consisting of multi-headed
rules that transform constraints into simpler ones until they are
solved. \chr\ defines both {\em simplification} of and {\em
  propagation} over user-defined constraints. Simplification replaces
constraints by simpler constraints while preserving logical
equivalence. Propagation adds new constraints, which are logically
redundant but may cause further simplifications.  Consider the
constraint \emph{min}, where $min(X,Y,Z)$ means that $Z$ is the
minimum of $X$ and $Y$. Then typical CHR rules for this constraint
are:
\begin{eqnarray*}
min(X,Y,Z), \  Y{\leq}X & \ \Leftrightarrow \ &  Z{=}Y, \  Y{\leq}X.\\
min(X,Y,Z), \  X{\leq}Y & \ \Leftrightarrow \ &   Z{=}X, \  X{\leq}Y.\\
min(X,Y,Z) & \ \Rightarrow \ & Z{\leq}X, \  Z{\leq}Y.
\end{eqnarray*} 
The first two rules are simplification rules, while the third one is a
propagation rule.  The first two rules correspond to the usual
definition of \emph{min}. The first rule states that $min(X,Y,Z), \;
Y{\leq}X$ can be replaced by $Z{=}Y, \; Y{\leq}X$. The second has an
analogous reading, and the third rule states that if we have
$min(X,Y,Z)$ then we can add $Z{\leq}X, \; Z{\leq}Y$ to the current
constraints.

If such rules are in general easy to read, in many cases it remains a
hard task to find them when one wants to write a constraint solver.
Thus, several methods have been proposed to automatically generate
rule-based solvers for constraints given their logical
specification~\cite{AptMonfroy99,RingeissenMonfroy99,AbdennadherRigotti00,AbdennadherRigotti01b}.
These approaches can help to find more easily interesting rules, and
it has also been shown in~\cite{AbdennadherRigotti01}
that the rules produced automatically can lead
to more efficient constraint reasoning than rules found by programmers.

In this paper, we propose a new method to generate CHR propagation and
simplification rules.  This work extends the previous rule-based
solver generation techniques described
in~\cite{AptMonfroy99,RingeissenMonfroy99,AbdennadherRigotti00,AbdennadherRigotti01b}.
It allows to obtain more general forms of rules, even for constraints
defined intensionally over infinite domains.

The intuitive principle of the generation is the following.  Consider
that a solver $S$ for some constraints, called {\em primitive
  constraints}, is already available. Other constraint predicates,
called {\em user-defined constraints}, are given and their semantics
is specified by mean of a constraint logic program $P$ (i.e., the
constraint predicates are defined by clauses of $P$).  Then we want to
obtain CHR propagation and simplification rules for the user-defined
constraints to extend the existing solver.  The basic idea of our
approach relies on the following observation: a rule of the form $C
\Rightarrow D$ is valid if the execution of the goal $C, \lnot(D)$
finitely fails with program $P$ and solver $S$.  For the execution of
such goals, we will use a tabled resolution for constraint logic
programming~\cite{CuiWarren00} that terminates more often
than execution based on SLD-like resolutions.

We present three algorithms that can be integrated to build an
environment to help developers to write CHR rule-based constraint
solvers.  Two of the algorithms focus on how to generate propagation
rules for constraints given their logical specification.  The first
algorithm generates only primitive propagation rules (i.e.,\ rules
with right hand side consisting of primitive constraints). The second
algorithm extends the first one to generate more general propagation
rules with right hand side consisting of both primitive and
user-defined constraints. We also show that a slight modification of
this algorithm allows to generate the so-called splitting rules (rule
having a disjunction in their right hand side) supported by the
extension of CHR called CHR$^\lor$~\cite{AbdennadherSchuetz98}.  The
third algorithm focuses on transforming propagation rules into
simplification rules to improve the time and space behavior of
constraint solving.

\paragraph{Organization of the paper.}

In Section~\ref{sec:primprop}, we present an algorithm to generate
primitive propagation rules.  In Section~\ref{sec:prop}, we describe
how to modify the algorithm to generate more general propagation
rules.  Section~\ref{sec:simp} presents a transformation method of
propagation rules into simplification rules. For clarity reasons, we
will use an abstract representation for the generated rules. In
Section~\ref{rule-simplification}, we present how these rules can be
encoded in \chr. We discuss related work in
Section~\ref{section-related-work}, and finally we conclude with a
summary and possibilities of further improvements.

\section{Generation of Primitive Propagation Rules}\label{sec:primprop}


\newcommand{\existsClosure}{{\tilde{\exists}}}
\newcommand{\primitivepropagminer}{{{\sc prim-miner}}}

We assume some familiarity with constraint logic
programming (CLP)~\cite{JaffarMaher94,MarriottStuckey98}.
We consider two
classes of constraints, {\em primitive constraints} and
{\em user-defined constraints}.  Primitive constraints are those
constraints defined by a constraint theory $CT$ and for which solvers
are already available. In the following, we do not expect that the solver for primitive constraints is complete. User-defined constraints are those
constraints defined by a constraint logic program $P$ and for which we
want to generate solvers.  We assume that the set of primitive
constraints is closed under negation, in the sense that the negation
of each primitive constraint must be also a primitive constraint,
e.g.\ $=$ and $\not=$ or $\leq$ and $>$. In the following, we denote
the negation of a primitive constraint $c$ by $not(c)$. 

In the rest of this paper, we use the following terminology.

\begin{definition}
  {\rm A \emph{constraint logic program} is a set of clauses of the
    form

\[h \leftarrow b_1, \ldots, b_n, c_1, \ldots, c_m\]

where $h, b_1, \ldots, b_n$ are user-defined constraints and $c_1,
\ldots, c_m$ are primitive constraints.  $h$ is called left hand side
of the clause. A \emph{goal} is a set of primitive and user-defined
constraints. An \emph{answer} is a set of primitive constraints.  The
logical semantics of a constraint logic program $P$ is its Clark's
completion and is denoted by $P^*$.  A user-defined constraint is
\emph{defined} in a constraint logic program if it occurs in the left
hand side of a clause.}
\end{definition}

\begin{definition}
  {\rm A \newterm{primitive propagation rule} is a rule of the form
    $C_1 \propag C_2$ or of the form $C_1 \propag false$, where $C_1$
    is a set of primitive and user-defined constraints, while $C_2$
    consists only of primitive constraints.  $C_1$ is called the
    \emph{left hand side} of the rule ({\em lhs}) and $C_2$ its
    \emph{right hand side} ({\em rhs}).  A rule of the form $C_1
    \propag \false$ is called \emph{failure rule}.  }
\end{definition}

In the following we use the notation $\existsClosure(\phi)$
to denote the existential closure of $\phi$ and
$\exists_{ -{\cal V}}(\phi)$ to
denote the existential closure of $\phi$ except for the variables in
the set ${\cal V}$.

\begin{definition}
  {\rm A primitive propagation rule $\{d_1, \ldots, d_n\} \propag
    \{c_1, \ldots, c_m\}$ is \newterm{valid} with respect to the
    constraint theory $\ctheory$ and the program $P$ if and only
    if $P^*, \ctheory \models \bigwedge_i d_i \rightarrow \exists_{
      -{\cal V}} (\bigwedge_j c_j)$, where ${\cal V}$ is the set of
    variables appearing in $\{d_1, \ldots, d_n\}$.  A failure rule
    $\{d_1, \ldots, d_n\} \propag \false$ is \newterm{valid} with
    respect to $\ctheory$ and $P$ if and only if $P^*, \ctheory
    \models \neg \existsClosure ( \bigwedge_i d_i)$.  }
\end{definition}

We now give an algorithm to generate valid primitive propagation
rules.

\subsection{The  \primitivepropagminer\ Algorithm}~\label{section-algo-primitive-miner}

The \primitivepropagminer\ algorithm takes as input the program $P$
defining the user-defined constraints.  To specify the syntactic form
of the rules, the algorithm needs also as input two sets of primitive
and user-defined constraints denoted by $\baselhs$ and $\candlhs$, and
a set containing only primitive constraints denoted by $\candrhs$.
The constraints occurring in $\baselhs$ are the common part that must
appear in the lhs of all rules, $\candlhs$ indicates candidate
constraints used in conjunction with $\baselhs$ to form the lhs, and
$\candrhs$ are the candidate constraints that may appear in the rhs.

Note that a syntactic analysis of the constraint logic program $P$ can
suggest functors and constraint predicates to be used to form
candidate constraints. 

The algorithm \primitivepropagminer\ is presented in
Figure~\ref{figure-algo-primitivepropagminer} and generates a set of
valid rules of the form $C \propag d$ where $d$ is a single
  primitive constraint. Note that rules with the same lhs, e.g. $C
  \propag d_1$, $C \propag d_2$, \ldots, $C \propag d_n$, can be
  replaced by the  single rule $C \propag d_1, d_2, \ldots,
  d_n$.


The basic idea of the algorithm relies on the following observation:
to be able to generate a failure rule of the form $C \propag false$,
we can simply check that the execution of the goal $C$ finitely fails.
Furthermore, while these rules are useful to detect inconsistencies,
it is in general more interesting to propagate earlier some
information that can be used for constraint solving, instead of
waiting until a conjunction of constraints becomes inconsistent.
Thus, for each possible lhs $C$ (i.e., each subset of $\baselhs \cup
\candlhs$) the algorithm distinguishes two cases:
%

\begin{enumerate}
\item \primitivepropagminer\ uses a CLP system
  to evaluate the goal $C$. If the goal finitely fails, then the
  failure rule $C \propag false$ is generated.
\item Otherwise the negation of each candidate constraint $d$ from
  $\candrhs$ is added in turn to $C$ and the goal $C \cup \{not(d)\}$) is
  evaluated. If the goal finitely fails, then the rule $C \propag
  \{d\}$ is generated.
\end{enumerate}

\begin{figure}[h]~\label{figure-algo-primitivepropagminer}
\hrulefill
\begin{center}
\begin{tabbing}

\kw{begin}\\
xxxx\=xxxx\=xxxx\=xxxx\=xxxx\=xxxx\=xxxx\= \kill
\\
\> \newruleset\ the resulting rule set is initialized to the empty set.\\

\> \Llhs\ is a list of all subsets of $\candlhs$,\\
\> in an order compatible with the subset partial ordering\\
\> (i.e., for all $C_1$ in \Llhs\ if $C_2$ is after $C_1$ in \Llhs\ then
$C_2 \not\subset C_1$).\\
\\
\> \kw{while} \Llhs\ is not empty \kw{do}\\

\>\>  Remove from \Llhs\ its first element denoted $C_\lhs$.\\

\>\>  \kw{if} the goal $(\baselhs \cup C_\lhs)$ fails\\
\>\>\ \ \    with respect to the constraint logic program $P$ \kw{then}\\

\>\>\>    add the failure rule $(\baselhs \cup C_\lhs \propag false)$ to
\newruleset\\
\>\>\>    and remove from \Llhs\ all supersets of $C_\lhs$.\\

\>\> \kw{else}\\
\>\>\> Let $rhs$ be initialized to the empty set.\\
\>\>\> \kw{for all} $d \in \candrhs$\\
\>\>\>\> \kw{if} the goal $(\baselhs \cup C_\lhs \cup \{not(d)\})$ fails\\
\>\>\>\>\ \ \  with respect to the constraint logic  program $P$ \kw{then}\\
\>\>\>\>\> add $d$ to the set $rhs$.\\
\>\>\>\> \kw{endif}\\
\>\>\> \kw{endfor}\\
\>\>\> \kw{if} $rhs$ is not empty \kw{then} add the rule
           $(\baselhs \cup C_\lhs \propag rhs)$ to \newruleset\ \kw{endif}\\
\>\> \kw{endif}\\

\>\kw{endwhile}\\
\\
\>output \newruleset.\\
\\
\kw{end}

\end{tabbing}

\end{center}

\hrulefill
\caption{The \primitivepropagminer\ Algorithm}
\end{figure}

In practice these goal evaluations are made using a bounded depth
resolution procedure to avoid non-termination of the whole generation
algorithm.

\smallskip

The algorithm \primitivepropagminer\ uses a basic ordering to prune the
search space and to avoid the generation of many uninteresting rules.
This pruning relies simply on the following observation.  If $C_1
\propag false$ is valid, then rules of the form $C_2 \propag false$,
where $C_1 \subset C_2$ are also valid but useless.  So the algorithm
considers first the smallest lhs with respect to set inclusion, and
when it finds a valid failure rule $C_1 \propag false$ it discards
from the lhs candidates any  $C_2$ that is superset of $C_1$.

At first glance, the procedure used to evaluate the goals issued by
the algorithm may be considered as a classical depth-first,
left-to-right CLP resolution.  However, we will show in
Section~\ref{section-interest-tabled} and Section~\ref{sec:prop} that
a tabled CLP resolution extends greatly the class of rules that can be
generated, by allowing termination of the evaluation in many
interesting cases.  Additionally, it should be noticed that the
execution on the underlying CLP system is not required to enumerate
all answers since \primitivepropagminer\ only performs a fail/succeed
test, and thus the CLP system can stop after a first answer has been
found.


\begin{example}~\label{ex:min}
  \rm Consider the following constraint logic program which implements
  the predicate \emph{min}.  $min(X,Y,Z)$ means that $Z$ is the
  minimum of $X$ and $Y$:
\begin{eqnarray*} 
min(X,Y,Z) & \ \leftarrow \ & X{\leq}Y, \ \ Z{=}X.\\
min(X,Y,Z) & \ \leftarrow \ & Y{\leq}X, \  \ Z{=}Y.
\end{eqnarray*} 
\noindent where $\leq$ and $=$ are primitive constraints with the usual
meaning as total order and syntactic equality.

The algorithm with the appropriate input generates (among others) the
rule
\begin{eqnarray*} 
min(X,Y,Z),  \ Y{\leq}X & \ \Rightarrow \ &  Z{=}Y.
\end{eqnarray*}
after having checked that the execution of the goal $min(X,Y,Z), \ 
Y{\leq}X, \ Z{\not=}Y$ fails by constructing the following derivation
tree:

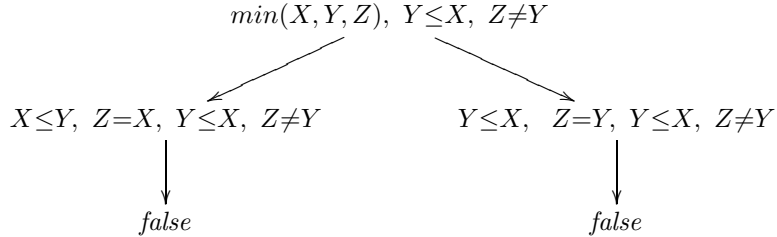
\[
\xymatrix@C-15ex{
&   min(X,Y,Z), \ Y{\leq}X, \ Z{\not=}Y \ar[dl]
\ar[dr] \\
X{\leq}Y, \  Z{=}X, \ Y{\leq}X, \ Z{\not=}Y \ar[d] &&  Y{\leq}X, \  \
Z{=}Y, \ Y{\leq}X, \ Z{\not=}Y \ar[d] \\
\false && \false
}
\]

Note that we assume that the constraint solver for $\leq$ and $=$ is
able to detect such inconsistencies.
\end{example}

\paragraph{Soundness and Completeness.} The \primitivepropagminer\  algorithm attempts to extract all valid primitive propagation rules of the form
$C_1 \propag \{d\}$ or $C_1 \propag \false$ such that $\baselhs
\subseteq C_1$, $C_1 \setminus \baselhs \subseteq \candlhs$, $d \in
\candrhs$, and there is no other more general failure rule (i.e., no
valid rule $C_2 \propag \false$ where $C_2 \subset C_1$).  In general,
the algorithm cannot be complete, since the evaluation of some goals
corresponding to valid rules may be non-terminating.  In fact, this
completeness can be achieved if more restricted classes of constraint
logic programs are used to give the semantics of user-defined
constraints and if the solver for the primitive constraints used by
the underlying CLP system is satisfaction complete.

The soundness of the algorithm (i.e., only valid rules are generated)
is guaranteed by the nice properties of standard CLP
schemes~\cite{JaffarMaher94} and tabled CLP schemes~\cite{CuiWarren00}.
An important practical aspect, is that even a partial resolution
procedure (e.g., bounded depth evaluation) or the use of an incomplete
solver by the CLP system, does not compromise the validity of the
rules generated.

\subsection{Advantage of Tabled Resolution for Rule Generation}~\label{section-interest-tabled}

Termination of the evaluation of (constraint) logic programs has
received a lot of attention. A very powerful and elegant approach
based on tabled resolution has been developed, first for logic
programming (e.g.,~\cite{Tamaki&Sato86,Warren92}) and further extended
in the context of CLP (e.g.,~\cite{CuiWarren00}).

The intuitive basic principle of tabled resolution is the following.
Each new subgoal $S$ is compared to the previous intermediate subgoals
(not necessarily in the same branch of the resolution tree).  If there
is a previous subgoal $I$ which is equivalent to $S$ or more general
than $S$, then no more unfolding is performed on $S$ and answers for
$S$ are selected among the answers of $I$. This process is repeated
for all subsequent computed answers that correspond to the subgoal
$I$.

The use of such technique has not been widely accepted since if this
leads to termination in many more cases than execution based on
SLD-resolution, this should be paid by some execution overhead in
general.  When using the algorithm \primitivepropagminer\ we can
accept a slight decrease of performance (since the solver is
constructed once) if this gives rise to an improvement of the
termination capability which enables the generation of additional
rules.  Thus, tabled CLP can find very interesting applications in
constraint solver synthesis.  This will be illustrated in the
following example.

\begin{example}
  {\rm Consider the well-known ternary \emph{append} predicate for
    lists, which holds if its third argument is a concatenation of the
    first and the second argument.  
\begin{eqnarray*}
append(X,Y,Z) & \ \leftarrow \ & X {=} [], \  Y {=} Z.\\
append(X,Y,Z) & \ \leftarrow \ & X {=} [H|X1], \ Z {=} [H|Z1], \ append(X1,Y,Z1). 
\end{eqnarray*}
Let the input parameters of \primitivepropagminer\ be
\begin{eqnarray*}
\baselhs & = & \{append(X,Y,Z)\} \\
\candlhs & = & \{X{=}[],Y{=}[],Z{=}[],X{=}Y,X{=}Z,Y{=}Z,X{\not=}Y,X{\not=}Z,Y{\not=}Z\}\\
\candrhs & = & \candlhs.
\end{eqnarray*}
Then the algorithm generates (among others) the
following primitive propagation rule:
\begin{eqnarray*}
append(X,Y,Z),\ Y{=}[] & \ \Rightarrow \ & X{=}Z
\end{eqnarray*}
by executing the goal $append(X,Y,Z), \ Y{=}[], \ X{\not=}Z$ with a
tabled CLP resolution.  For a classical CLP scheme the resolution tree
will be infinite, while in case of a tabled resolution it can be
sketched as follows:
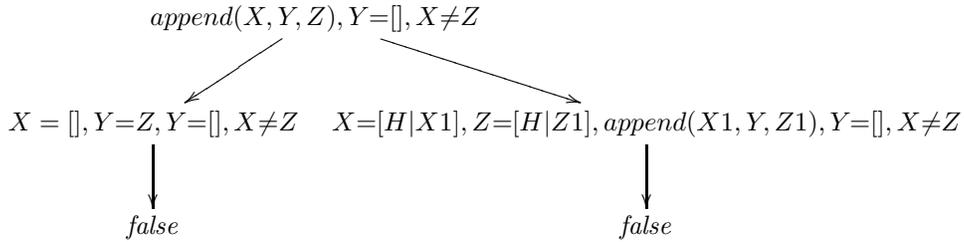
\[
\xymatrix@C-20ex{
&  append(X,Y,Z), Y{=}[], X{\not=}Z \ar[dl]
\ar[dr] \\
X=[], Y{=}Z, Y{=}[], X{\not=}Z  \ar[d] &&  X{=}[H|X1], Z{=}[H|Z1], 
append(X1,Y,Z1), Y{=}[],  X{\not=}Z \ar[d] \\
\false && \false
}
\]

The initial goal $G_1 = (append(X,Y,Z), \ Y{=}[], \ X{\not=}Z)$ is
more general than the subgoal $G_2 = (X{=}[H|X1], \ Z{=}[H|Z1], \ 
append(X1,Y,Z1), \ Y{=}[], \ X{\not=}Z$), in the sense that
$append(A,B,C), \ D{=}[E|A], \ F{=}[E|C], \ B{=}[], \ D{\not=}F$
entails $append(A,B,$ $C),
 \ B{=}[], \ A{\not=}C$.  So no unfolding is
made on $G_2$, and the process waits for answers of $G_1$ to compute
answers of $G_2$.  Since $G_1$ has no further possibility of having
answers, then $G_2$ fails and thus $G_1$ also fails.  We refer the
reader to~\cite{CuiWarren00} for a more detailed
presentation of goal evaluation using a tabled CLP resolution.

Using this kind of resolution \primitivepropagminer\ is also able to
produce rules such as:
\begin{eqnarray*}
append(X,Y,Z),\ X{=}Z \ & \Rightarrow & \ Y{=}[]. \\
append(X,Y,Z),\ Y{\not=}[] \ & \Rightarrow & \ X{\not=}Z.\\
append(X,Y,Z),\ X{\not=}[] \ & \Rightarrow & \ Z{\not=}[]. 
\end{eqnarray*}
}
\end{example}

We have run all examples presented in this paper, using our own
implementation of tabled CLP resolution according to the description
of~\cite{CuiWarren00}. The running prototype is implemented in SICStus
Prolog 3.7.1.

\subsection{Selection of Interesting Rules and Performance of Generation}

\paragraph{Interesting rules.}
If we use the algorithm \primitivepropagminer\ as presented in
Figure~\ref{figure-algo-primitivepropagminer}, we obtain a set of
rules that is highly redundant and thus not suitable neither for
human-reading nor for its direct use as an executable constraint
solver.  This problem encountered in propagation rule generation
has already been pointed out
in~\cite{AbdennadherRigotti00}.
It is taken into account by applying the following simplification
technique on the set of rules produced by \primitivepropagminer:

\begin{itemize}

\item The rules generated by \primitivepropagminer\ are
  ordered in a list $\Llhs$ using any
  total ordering on the rule lhs compatible with the
  $\theta$-subsumption ordering~\cite{plotkin70} (i.e., a rule having
  a more general lhs is placed before a rule with a more specialized
  lhs).

\item Let $S$ be a set of rules initialized to the empty set.  For
  each rule $C_1 \propag C_2$ in $\Llhs$ (taken according to the list
  ordering) the constraint $C_2$ is simplified to an equivalent
  constraint $C_{simp}$ by the already known solver
  for \candrhs\ and by the rules in $S$.
  If $C_{simp}$ is empty then the rule can be discarded,
  else add the rule $C_1 \propag C_{simp}$ to $S$.
 
\item Output the set $S$ containing the simplified set of rules.
\end{itemize}

For example, using this process the set of rules
\[\{
p(X) \propag r(X), \ \ \ \ p(X), \; q(X) \propag r(X), \ \ \ \ s(X,Y) \propag
X{=}Y, \; X{=}a, \; Y{=}a \}\] is simplified to \[ \{ p(X) \propag
r(X),\ \ \ \ s(X,Y) \propag X{=}a, \; Y{=}a \}.\]

For clarity reasons this simplification step is presented separately from
the algorithm \primitivepropagminer, but it can be incorporated in
\primitivepropagminer\ and performed during the generation of the
rules.  So, if the algorithm \primitivepropagminer\ is implemented on
a flexible platform (e.g., SICStus Prolog with CHR support), when a
valid rule is extracted it can be immediately simplified with respect
to the already generated rules. Then, if it is not redundant it can be
incorporated at runtime and thus be used actively to speed up further
resolution and constraint solving steps called by the generation
algorithm itself.

\paragraph{Other performance issues.}
It is likely that the same constraints and their negations are
candidates for both the lhs and rhs of the rules. In this case the two
following optimizations can be used:

\begin{enumerate}
\item If a rule $C \propag \{d\}$ is generated and $not(d) \in \candlhs$ then there is no need to generate the rule $C \cup \{not(d)\} \propag false$ since this rule is trivially redundant.
\item
If a rule $C_1 \propag \{d\}$ is generated and $d$ is also in $\candlhs$ then
there is no need to consider any $C_2$ such that $C_2$ is a superset
of $C_1$ containing $d$ to form the lhs of another rule.
\item If a rule $C \cup \{d_1\} \propag \{d_2\}$ is produced
using the evaluation of the goal $C \cup \{d_1,\ not(d_2)\}$
(which fails), and  $not(d_2) \in \candlhs$ and $not(d_1) \in
\candrhs$ then the same goal evaluation is also needed to produce the rule
$C \cup \{not(d_2)\} \propag \{not(d_1)\}$. Thus, we can avoid for such constraints
this kind of repeated goal evaluations.

For example, the execution of the goal $min(X,Y,Z), \ Y{\leq}X, \ 
Z{\not=}Y$ may lead to the generation of the following rules:
\begin{eqnarray*} 
min(X,Y,Z), \ Y{\leq}X &\ \Rightarrow\ & Z{=}Y .\\
min(X,Y,Z), \ Z{\not=}Y &\ \Rightarrow\ & Y{>}X.
\end{eqnarray*} 

\end{enumerate}

\subsection{Generation of Primitive Splitting Rules}

Splitting rules have been shown to be interesting in constraint
solving~\cite{Apt98}, since they can be used to detect early
alternative labeling cases or alternative solution sets.
These rules are handled by CHR$^\lor$ an extension of
CHR proposed in~\cite{AbdennadherSchuetz98}.
Such rules
that can be generated by the extension proposed in this section are
for example:
\begin{eqnarray*} 
and(X,Y,Z),\ \ Z{=}0 &\ \Rightarrow\ & X{=}0\ \  \vee \ \ Y{=}0.\\
min(X,Y,Z) &\ \Rightarrow\ & X{=}Z\ \ \vee \ \ Y{=}Z.
\end{eqnarray*} 

where $and(X,Y,Z)$ means that $Z$ is the Boolean conjunction of $X$
and $Y$, defined by the facts $and(0,0,0), and(1,0,0), and(0,1,0)$,
$and(1,1,1)$; and where $min(X,Y,Z)$ is defined by the constraint
logic program of Example~\ref{ex:min}.

In the following, we restrict ourself to the generation of
\emph{primitive splitting rules}.

\begin{definition}
  {\rm A {\em primitive splitting rule} is a rule of the form $C
    \propag d_1 \vee d_2$, where $d_1, d_2$ are primitive constraints
    and $C$ is a set of primitive and user-defined constraints.
    $\vee$ is interpreted like the standard disjunction, and primitive
    splitting rules have a straightforward associated semantics.  }
\end{definition}

The modification of the basic \primitivepropagminer\ algorithm is as
follows.  For all $C_\lhs$, it must also consider each different set
$\{d_1,d_2\} \subseteq C_\rhs$ with $d_1 \not= d_2$ and check if the
goal $(\baselhs \cup C_\lhs \cup \{not(d_1), not(d_2)\})$ fails with
respect to the constraint logic program $P$. If this is the case, it
simply adds to \newruleset\ the primitive splitting rule $(\baselhs
\cup C_\lhs \propag d_1 \vee d_2)$.  

For example, the rule $min(X,Y,Z) \Rightarrow X{=}Z\ \vee \ Y{=}Z$ can
be obtained by running the goal $min(X,Y,Z), \ X{\not=}Z, \ Y{\not=}Z$
and checking that its execution finitely fails.

Here again, the soundness of the generation relies on the properties
of the underlying resolution used.

In practice, a huge number of splitting rules are not interesting
because they are redundant with respect to some primitive propagation
rules.  So, to remove these uninteresting rules, the generation should
preferably be done, by first using \primitivepropagminer\ to obtain
only primitive propagation rules, and then using the modified version
of the algorithm to extract primitive splitting rules. Thus in this
second step, redundant rules, with respect to the set resulting from
the first step, can be discarded on-the-fly.  For example the rule
$a,b,c\ \Rightarrow\ d \vee e$ will be removed if we have already
generated the rule $a,b\ \Rightarrow\ d$.  Moreover, in such trivial
redundancy cases, the test of the validity of the rule $a,b,c\ 
\Rightarrow\ d \vee e$ can itself be avoided, and more generally the
test of any rule of the form $C \Rightarrow d \vee f$ and $C
\Rightarrow f \vee d$, where $\{a,b\} \subseteq C$ and $f \in \candrhs$.

\section{Generation of More General Propagation Rules}\label{sec:prop}

In this section, we modify the algorithm presented in
Section~\ref{sec:primprop} to handle a broader class of rules called
{\em general propagation rules}, that encompasses the primitive
propagation rules, and that can even represent recursive rules over
user-defined constraints (i.e., rules where the same user-defined
constraint predicate appears in the lhs and rhs).

\begin{definition}
  {\rm A \newterm{general propagation rule} is a failure rule or a
    rule of the form $C_1 \propag C_2$, where $C_1$ and $C_2$ are sets
    of primitive and user-defined constraints.}
\end{definition}

The notion of validity defined for primitive propagation rules also
applies to this kind of rules.

In the \primitivepropagminer\ algorithm, the validity test of a
primitive propagation rule $\baselhs \cup C_\lhs \propag \{d\}$ is
performed by checking that the goal $\baselhs \cup C_\lhs \cup
\{not(d)\}$ fails.  For general propagation rules, $d$ is no longer a
primitive constraint but may be defined by a constraint logic program.
In this case, the evaluation should be done using a more general
resolution procedure to handle negated subgoals. However, to avoid the
well known problems related to the presence of negation we can simply
use a different validity test based on the following theorem.

\begin{theorem}
Let $C_1 \Rightarrow C_2$ be a general propagation rule
and ${\cal V}$ be the variables occurring in $C_1$.
Let $S_1$ be the set of answers $\{a_1,\ldots, a_n\}$ to the goal
$C_1$, and $S_2$ be the set of answers $\{b_1,\ldots, b_m\}$ to
the goal $C_1 \cup C_2$.
Then  the rule $C_1 \Rightarrow C_2$ is valid if
$P^*, \ctheory \models
\lnot(\exists_{ -{\cal V}} ((a_1 \vee \ldots \vee a_n) \wedge
\neg \exists_{ -{\cal V}}(b_1 \vee
\ldots \vee b_m)))$. 
\end{theorem}

This straightforward property comes from the completeness result of standard CLP
schemes~\cite{JaffarMaher94} which ensures that if a goal $G$ has a
finite computation tree, with answers $c_1,\ldots,c_n$ then $P^*,
\ctheory \models G \leftrightarrow \exists_{ -{\cal V}_g}(c_1 \vee
\ldots \vee c_n)$, where ${\cal V}_g$ is the set of variables appearing
in $G$.

So, the modification proposed in this section consists simply in the
replacement in algorithm \primitivepropagminer\ of the call to the
goal $\baselhs \cup C_\lhs \cup \{not(d)\}$ to check the validity of
the rule $\baselhs\ \cup \ C_\lhs \propag \{d\}$, by the following
steps.

\begin{itemize}
\item First, collect the set of answers $\{a_1,\ldots, a_n\}$ to the
  goal $\baselhs \cup C_\lhs$.
  
\item Then, collect the set of answers $\{b_1,\ldots, b_m\}$ to the
  goal $\baselhs  \cup  C_\lhs  \cup  \{d\}$.
  
\item In each answer $a_i$ (resp. $b_i$), rename with a fresh variable
  any variable that is not in $\baselhs \cup C_\lhs$ (resp.
  $\baselhs \cup C_\lhs \cup \{d\}$).

\item  Finally,
  perform a satisfiability test of $(a_1 \vee \ldots \vee a_n) \wedge
  \neg (b_1 \vee \ldots \vee b_m)$.  
  
\item If this test fails then the rule $\baselhs \cup C_\lhs \propag
  \{d\}$ is valid.  
\end{itemize}

Since answers only contain primitive constraints and since the set of
primitive constraints is closed under negation, then we can perform
the satisfiability test by rewriting $(a_1 \vee \ldots \vee a_n)
\wedge \neg (b_1 \vee \ldots \vee b_m)$ into an equivalent
disjunctive normal form, and then use the solver for primitive
constraints on each sub-conjunctions.

It should be noticed that in cases where the evaluation of one of the
two goals does not terminate before the bound of resolution depth is
reached then the propagation rule $\baselhs \cup C_\lhs \propag \{d\}$
is not considered as valid and the next rule is processed.

\begin{example}
  {\rm Consider the following user-defined Boolean constraints:
    $neg(X,Y)$ imposing that $Y$ is the Boolean complement of $X$ and
    $xor(X,Y,Z)$ stating that $Z$ is the result of the exclusive
    Boolean disjunction of $X$ and $Y$.  These two constraints are
    defined by a straightforward constraint logic program.  Among
    others, the modified algorithm presented above can generate the
    following rules:

\begin{eqnarray*} 
xor(X,Y,Z), \ Z{=}1 & \Rightarrow & neg(X,Y). \\
xor(X,Y,Z), \ Y{=}1  & \Rightarrow & neg(X,Z).\\
xor(X,Y,Z), \ X{=}1  & \Rightarrow & neg(Y,Z). 
\end{eqnarray*}

For example, to test the validity of the first rule, the process is
the following.  First, the answers $A_1 := X{=}1 \wedge Y{=}0 \wedge
Z{=}1$ and $A_2 := X{=}0 \wedge Y{=}1 \wedge Z{=}1$ to the goal
$xor(X,Y,Z), \ Z{=}1$ are computed. Then
for the goal $xor(X,Y,Z), \ Z{=}1, \ neg(X,Y)$
the same answers are collected.
Finally
the satisfiability test of $(A_1 \vee A_2) \wedge \lnot (A_1 \vee
A_2)$ fails and thus establishes the validity of the rule.

One other general rule that can be generated using the modified
algorithm presented is for example:
\begin{eqnarray*}
and(X,Y,Z) & \ \Rightarrow \ & min(X,Y,Z).
\end{eqnarray*}
Furthermore, rules representing symmetries can be automatically
detected using the modified algorithm. For example, the rules
\begin{eqnarray*}
min(X,Y,Z) & \ \Rightarrow \ & min(Y,X,Z).\\
xor(X,Y,Z) & \ \Rightarrow \ & xor(Y,X,Z).
\end{eqnarray*}
expressing the symmetry of the \emph{min} and the \emph{xor}
constraints with respect to the first and second arguments can be
generated.  In~\cite{AbdennadherRigotti01b}, it has been shown that
these rules are very useful to reduce the size of a set of propagation
rules since many rules become redundant when we know such symmetries.}
\end{example}

However, it must be pointed out that if the test presented in this
section allows us to handle a syntactically wider class of rules, it
relies on different goal calls than the test of
Section~\ref{section-algo-primitive-miner}.  So when testing the
validity of a primitive propagation rule one of the techniques may
lead to terminating evaluation while the other one may not.  Thus in
the case of primitive propagation rule it is preferable not to replace
one test by the other, but to use both in a complementary way (run one
of them, and if it reaches the bound of resolution depth then apply
the other).

\section{Generation of Simplification Rules}\label{sec:simp}

Since a propagation rule does not rewrite constraints but adds new
ones, the constraint store may contain superfluous information.
Constraints can be removed from the constraint store using
\emph{simplification rules}.


\begin{definition}
  {\rm A \emph{simplification rule} is a rule of the form $C_1
    \Leftrightarrow C_2$, where $C_1$ and $C_2$ are sets of primitive
    and user-defined constraints.

}
\end{definition}

In this section, we show how the rule validity test used in
Section~\ref{sec:prop} can be applied to transform some propagation rules
into simplification rules.  For a
valid propagation rule of the form $C \Rightarrow D$, we try to find a
proper subset $E$ of $C$ such that $D \cup E \Rightarrow C$ is valid
too. If such $E$ can be found, the propagation rule $C \Rightarrow D$
can be transformed into a simplification rule of the form $C
\Leftrightarrow D \cup E$.

To simplify the presentation, we present an algorithm to transform
(when possible) propagation rules into simplification rules
independently from the algorithm presented in Section~\ref{sec:prop}.
Note that the algorithm for the generation of propagation rules can be
slightly modified to incorporate this step and to directly generate
simplification rules.

The algorithm is given in Figure~\ref{figure-algo-simpminer} and takes
as input the set of generated propagation rules and the common part
that must appear in the lhs of all rules, i.e.\ $\baselhs$.

\begin{figure}[ht]
\hrulefill

\begin{center}
\begin{tabbing}

\kw{begin}\\
xxxx\=xxxx\=xxxx\=xxxx\=xxxx\=xxxx\= \kill
\\
\>$P' := P$\\
\> \kw{for} each propagation rule of the form  $C
\Rightarrow D$ in $P$ \kw{do}\\
\>\>Find a proper subset $E$  of $C$  such that $\baselhs \not \subseteq
E$ and \\
\>\> $D \cup E \Rightarrow C$ is valid (using the validity test of
Section~\ref{sec:prop})\\
\>\> \kw{If} $E$ exists \kw{then}\\
\>\>\> $P' := (P' \backslash \{C \Rightarrow D\}) \cup \{C \Leftrightarrow D \cup E\}$\\
\>\> \kw{endif}\\
\> \kw{endfor}\\
\\
\>output $P'$\\

\kw{end}

\end{tabbing}

\end{center}

\hrulefill
\caption{The Transformation Algorithm  \label{figure-algo-simpminer}}
\end{figure}

To achieve a form of minimality based on the number of constraints, we
generate simplification rules that will remove the greatest number of
constraints. So, when we try to transform a propagation rule into a
simplification rule of the form $C \Leftrightarrow D \cup E$ we choose
the smallest set $E$ (with respect to the number of atomic constraints
in $E$) for which the condition holds.  If such a $E$ is not unique,
we choose any one among the smallest. The condition $\baselhs \not
\subseteq E$ is needed to be able to transform the propagation rules
into simplification rules that rewrite constraints to \emph{simpler}
ones (primitive constraints if possible), as shown in the following
example.

\begin{example}
  {\rm Consider the following propagation rule $R$ generated for the
    \emph{append} constraint with $\baselhs = \{append(X,Y,Z)\}$:
\begin{eqnarray*} 
append(X,Y,Z),\ X{=}[] & \ \Rightarrow \ &  Y{=}Z.
\end{eqnarray*}  
%
The algorithm tries the following transformations:

\begin{enumerate}
  
\item First, it checks if rule $R$ can be transformed into the
  simplification rule $append(X,Y,Z),\ X{=}[] \Leftrightarrow Y{=}Z.$
  This is done by testing whether the rule $Y{=}Z \Rightarrow
  append(X,Y,Z),\ X{=}[]$ is valid. But, this is not the case and thus
  the transformation is not possible.
  
\item Next, the algorithm finds out that the rule $Y{=}Z, \ X{=}[]
  \Rightarrow append(X,Y,Z)$ is a valid rule and then the propagation
  rule $R$ is transformed into the simplification rule
  $append(X,Y,Z),\ X{=}[] \Leftrightarrow X{=}[], \ Y{=}Z.$
\end{enumerate}

Note that the propagation rule $Y{=}Z, \ append(X,Y,Z) \Rightarrow
X{=}[]$ is also valid but transforming rule $R$ into
\begin{eqnarray*} 
append(X,Y,Z),\ X{=}[]  \ & \Leftrightarrow & \  append(X,Y,Z), \ Y{=}Z.
\end{eqnarray*} 
will lead to a simplification rule which is uninteresting for
constraint solving, i.e.\ using this rule the \emph{append} constraint
cannot be simplified and remains in the constraint store.  The
algorithm disables such transformation by checking the
condition that all constraints of $\baselhs$  are not shifted to the
right hand side of the rule ($\baselhs \not \subseteq E$).}
\end{example}

\section{Implementation of the Generated Rules in CHR}\label{rule-simplification}

The generated rules may contain constraints that are built-in
constraints for the CHR system.  To have a running CHR solver, these
constraints have to be encoded in a specific way.  First, equality
constraints appearing in the left hand side of a rule are propagated
all over the constraints in its left and right hand side. Then the
resulting constraints are simplified. This can be performed as
follows. In turn each equality constraint appearing in the lhs is
removed and transformed in a substitution that is applied to the lhs
and the rhs. Then the next equality constraint is processed. For
example, the simplification rule $and(X,Y,Z), \; Z{=}1 \Leftrightarrow
X{=}1, \; Y{=}1, \; Z{=}1$ will be transformed into $and(X,Y,1)
\Leftrightarrow X{=}1, \; Y{=}1$.  Secondly, for other built-in
constraints the transformation leads to \chr\ rules containing a
guard~\cite{Fru98}. For example, if $\leq$ is a built-in constraint of
the CHR system then the rule $min(X,Y,Z), \ X{\leq}Y \ \Leftrightarrow
\ Z{=}X, \ X{\leq}Y$ is transformed into the guarded \chr\ rule
$min(X,Y,Z) \ \Leftrightarrow \ X{\leq}Y \; | \; Z{=}X$.

\section{Related Work}~\label{section-related-work}

In~\cite{AbdennadherRigotti01b}, a method has been proposed to
generate propagation rules from the intentional definition of the
constraint predicates (eventually over infinite domains) given by
mean of a constraint logic program.  It extended previous
work~\cite{AptMonfroy99,RingeissenMonfroy99,AbdennadherRigotti00}
where different methods dedicated to the generation of propagation rules
for constraints defined extensionally over finite domains have been
proposed. Compared to the work described in~\cite{AbdennadherRigotti01b}
the approach presented in this paper has several advantages:

\begin{itemize}
\item It enables user-defined constraints
to occur in the right hand side of rules, while this is not handled
by~\cite{AbdennadherRigotti01b}.
As a by-product rules representing symmetries as
\begin{eqnarray*}
min(X,Y,Z) & \ \Rightarrow \ & min(Y,X,Z).
\end{eqnarray*}
can then be automatically detected.

\item It allows the generation of splitting rules like
\begin{eqnarray*}
append(X,Y,Z), \ Z{=}[A] & \ \Rightarrow \ &  X{=}[A] \  \lor \   Y{=}[A]
\end{eqnarray*} 
supported by
the extension of CHR called CHR$^\lor$~\cite{AbdennadherSchuetz98}.
These rules have been shown to be interesting in constraint
solving~\cite{Apt98}, but have not been considered
in~\cite{AptMonfroy99,RingeissenMonfroy99,AbdennadherRigotti00,AbdennadherRigotti01b}.

\item Even if we restrict our attention to the class of
rules handled in~\cite{AbdennadherRigotti01b}, the approach presented
in this paper leads to a more expressive set of rules. For example, the rule
\begin{eqnarray*}
append(X,Y,Z),\ Y{=}[] & \ \Rightarrow \ &  X{=}Z.
\end{eqnarray*} 
cannot be generated by the approach proposed
in~\cite{AbdennadherRigotti01b} while the algorithm described
in Section~\ref{sec:primprop}
is able to obtain it by executing the goal
$append(X,Y,Z),\ Y{=}[], X{\not=}Z$ with a tabled resolution for CLP.

\item Additionally, it needs less information about the semantics of the primitive
constraints. For example it generates the rule
\begin{eqnarray*} 
min(X,Y,Z) & \ \Rightarrow \ & Z{\leq}X, \ Z{\leq}Y.
\end{eqnarray*}
simply by calling the goals $min(X,Y,Z), \ Z{>}X$ and
$min(X,Y,Z), \ Z{>}Y$, and checking that their executions fail.
While in this case, the algorithm presented in~\cite{AbdennadherRigotti01b}
requires some extra information concerning the semantics of $\leq$
(information that would be provided in general by the user).
\end{itemize}

In this paper, we also described a method to transform propagation rules
into simplification rules.
It has been shown in~\cite{AbdennadherRigotti01} that this transformation
has a very important impact on the efficiency of the solver produced.
In~\cite{AbdennadherRigotti01} the propagation rules are modified to obtain
(when possible) simplification rules using a technique
based on a confluence notion. This is a syntactical criterion that works when we
have at hand the whole set of rules defining the constraint. Thus it
cannot be applied safely if only a part of the propagation rules have
been generated.  It also requires a termination test for rule-based
programs consisting of propagation and simplification rules, and this
test is undecidable for some classes of programs. The new transformation method
presented in this paper avoids these two restrictions.

\medskip

The generation of propagation and simplification rules is also related in some
aspects to \emph{Generalized Constraint
  Propagation}~\cite{ProvostWallace93}, \emph{Constructive
  Disjunction}~\cite{VanHentenryck+98,WuertzMueller:96a}, and
\emph{Inductive Logic Programming}~\cite{muggletonderaedt94}
as briefly discussed in~\cite{AbdennadherRigotti01b}.
However, it should be pointed out that to our knowledge these works have not
been used for the generation of constraint solvers.


\section{Conclusion and Future  Work}~\label{sec:conclusion}

In this paper, we have presented three algorithms that
can be integrated to build an environment
to help solver developers when writing CHR programs.

The approach described allows the generation of CHR propagation and
simplification rules.  It can be applied on constraints defined over
finite and infinite domains by mean of a constraint logic program.
Moreover, it enables the developer to search for rules having such
user-defined constraints in both their left and right hand sides.

We have also shown that compared to the algorithms described
in~\cite{AptMonfroy99,RingeissenMonfroy99,AbdennadherRigotti00,AbdennadherRigotti01b}
to generate rule-based constraint solvers,
this approach is able to generate more expressive rules (including
recursive and splitting rules).

One interesting direction for future work is to investigate the
integration of constructive negation
(e.g.,~\cite{IC::Stuckey1995}) in tabled resolution for CLP to
generate constraint solvers, in order to check the validity of the
propagation and simplification rules in more general cases.  Another
complementary aspect is the completeness of the solvers generated.  It
is clear that in general this property cannot be guaranteed, but in
some cases it should be possible to check it, or at least to
characterize the kind of consistency the solver can ensure.

\medskip



\end{document}